# MetaCache: Efficient Metadata Caching in Linux file system

Tanmay Agrawal, Ashay Shirwadkar, Pratik Gaikar, Kushagra Verma

Department of Information Technology
Maharashtra Institute of Technology
Pune- 411038, India
{tanmay0208, ashayshirwadkar12, pratikgaikar2008, verma.kush16} @gmail.com

*Abstract*— **Nowadays, Linux file systems have to manage millions of tiny files for different applications, and face with higher metadata operations. So how to provide such high metadata performance with such enormous number of files and large scale directories is a big challenge for Linux file system. We viewed that metadata lookup operations dominate metadata workload and incur low metadata performance. In this paper, we present a metadata cache to accelerate metadata access for Linux file system. Through this optimization, the Linux file system (such as EXT2, EXT4, BTRFS, etc.) can gain improvement in read rates as well as write rates.**

*Keywords*— *Metadata, NoSQL, Kernel, Castle, LSM tree.*

I. INTRODUCTION

With the amount of data increasing at an alarming rate, the performance of metadata plays an important role in achieving high I/O scalability and throughput of any Linux system. Linux file system performance is mostly dominated by metadata access. Data and metadata are stored as block in Linux system. Limited metadata update outcomes in full block write or read which amplifies Hard-disk input/output largely. Accessing files with state-of-the-art file systems typically results in multiple seeks. At first the location of the metadata (inode) has to be located using directory information, which results – if not cached – in multiple seeks. After the metadata is read, another head movement is required to read the actual file data on physical storage. In a scenario, where a huge number of small files need to be accessed, these head movements can slow access times tremendously because of seek time. Research work has reported that only 21% of requests are file Input/output whereas metadata operations are more than 50% [2]. Therefore, improving metadata I/O operation is desirable.

Linux Kernel should accept techniques from modern key-value stores for small files and metadata, because mention systems are "thin" enough to produce the performance levels needed by the systems. To inspire our assertion, we present an idea to display that for workloads influenced by small files and metadata, it is viable to improve the performance and scalability of the most modern local file systems(ext4,btrfs,etc) in Linux. It is done by adding an interposed MetaCache layer that represents the metadata in a LSM or Castle key-value store.

II. IMPORTANT TERMS

Even in the age of big data, most things in file system are limited. Surely, the number of small files or metadata in Linux local file system is to soon attain and exceed billions, a noted issue for both most and the largest Linux local file systems.

A. *Metadata*

Metadata is defined as data about data or information about information. For example, in a notepad, the sequence of alphabets that notepad contains is the actual data. But, the notepad has a creation date, a name, user ID of owner, a position on the storage area, protection, etc. All this data is metadata [7].

B. *Inode*

The Inode is an Data structure that contains metadata information (Inode number, Access Control List, Extended attribute ,Direct/indirect disk blocks ,Number of blocks, File generation number, File size, File type, Group, Number of links, Owner, Permissions) about files in UNIX file systems. Each file has an inode which is determined by an i-number (inode number) in the UNIX file system [8].

C. *Virtual File System (VFS)*

Virtual file system (VFS) provides an interface between User level system call and Linux file system (like NFS, Ext2, Ext4, BTRFS, etc.). The virtual file system acts as an abstraction layer that provides applications access to many types of file systems and network storage devices. That is why VFS is also called as virtual file system switch. The VFS also maintains Inode and Directory Cache so that inodes for regularly accessed directories can be located easily [9].

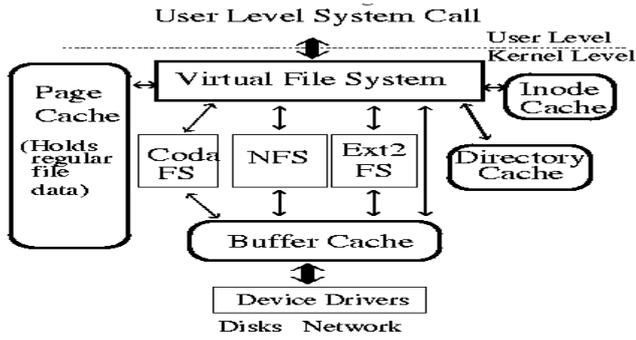

Fig. 1. The Linux Kernel's VFS[9]

*D. EXT4*

The Fourth Extended File system is design as a powerful file system for UNIX (Linux). It is an extension of Ext3 file system. Ext4 is also called a journaling filesystem. The Ext4 filesystem is built on the ground that the data held in files are kept in data blocks.

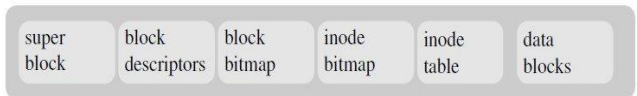

Fig. 2. Structure of an Ext4 Block Group

Fig.2 shows the structure of an ext4 block group. Superblocks contains a brief of the basic shape and size of the file system. The block descriptors contain pointers to the inode, block bitmap, inode Bitmap as well as the inode tables of all block groups. The block and inode bitmaps represent the block number of allocation bitmap which is used during block and deallocation. The inode table stores the actual inodes of files stored in this block. Ext4 inode contains all the metadata.

*E. LSM Tree*

LSM-tree stands for Log-Structured Merge-Tree. It is a data structure having performance characteristics that make it in interesting for giving indexed access to files with high insert volume. In a easy explanation of an LSM tree, a memory cache delays writing changed and new entries until it has a convincing magnitude of changes to book on Hard-disk. Delay writes in LSM-tree are made enduring by constantly booking changed and new entries in a write ahead log, which is pushed to Hard-disk regularly and asynchronously.

LSM tree contains 2 tree-like structures, M0 and M1. M0 component is smaller and entirely resident on RAM, whereas M1 component is resident on Hard-disk. New records are inserted onto the RAM-resident M0 component. If the insertion causes the M0 component to exceed a threshold size, a segment of entries is removed from M0 and merged into M1 on Hard-disk [10].

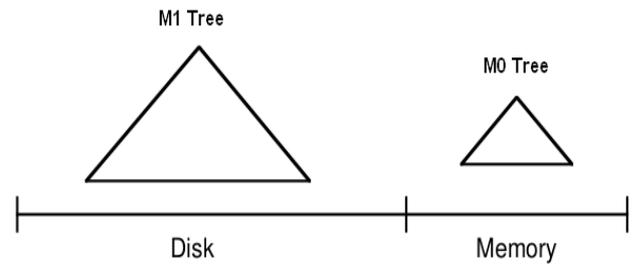

Fig. 3. Two components of LSM-tree [10]

### III. METACACHE

Metacache is a abstract cache which gives inode of corresponding file given an inode number. This cache is implemented just below VFS.
There are two cases while handling normal inode request,
- If VFS has already cached the inode in I-cache, then a corresponding request is served by VFS only
- If VFS hasn't cached the inode, then the request is forwarded to Hard Disk which surely holds the inode.

Metacache is implemented in such a fashion that while booting the computer, it loads all the metadata into RAM and then serves the next inode request.

Metacache can be implemented by either LSM tree or Acunu castle database. We have implemented using LSM tree, but we will discuss both.

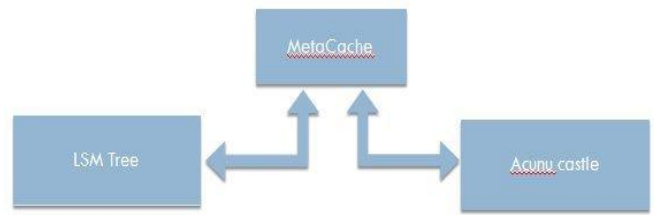

Fig.4. the Design suggested one is LSM and other is Castle

As Metacache uses LSM tree, it smartly stores recently used metadata into RAM (MemTable) and the remaining data is stored on hard disk. LSM Tree manages on-disk data as multiple large sorted arrays, known as SSTables. It maintains data in log-structured way. When inserting or updating data, first it updates data on MemTable, provided MemTable has corresponding entry, these changes are reflected back to SSTable.

MemTable is fixed in size, so next overflowing write request will dump the data into SSTable. When querying an element, request will have to traverse through all the SSTables which is an overhead. So every SSTable maintains Bloom Filter to reduce false positive lookups. Bloom Filter works as bitmap, showing weather entry is valid or not. Compaction is process of improving read query speed by using merge-sort on the list of SSTables.

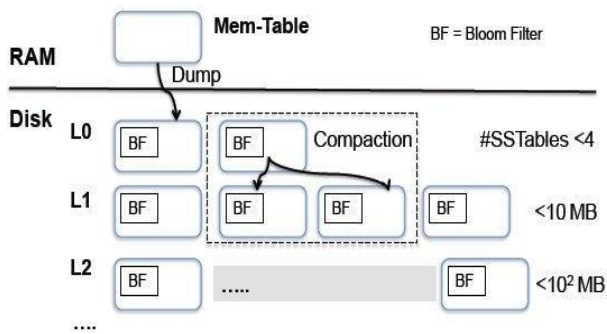

Fig 5: LSM Tree Architecture

MetaCache design is inspired by Combination of two ideas from prior work.
- **Co-locate metadata and small files**
  MetaCache is implemented in Ext4 filesystem. This filesystem has inbuilt feature of co-locating metadata (inode) and actual data in same block. This ultimately reduces random reads to hard disk.
- **Buffer writes in memory, store LSM tree**
  The data is not written to disk directly, LSM logs the data and uses delayed write which reduces random writes to hard disk.

Let's discuss an implementation using Acunu Castle. Acunu is database solution designed for handling bigdata. Castle is open source project which is the storage component of Acunu. As with LSM Tree Castle also stores data in key-value pairs. Castle supports snapshots of data; due to which it's possible to duplicate as well as taking backup is done efficiently. Snapshots are merely difference between two datasets, due to which memory wastage is avoided. Apart from above castle also provides caching and indexing of data as well as redundancy and disk layout management.

Castle is divided into two modules – a Linux kernel module, handling storage functionality and a userspace control daemon. But here we are only interested in Linux kernel module. We have to modify the code such that, VFS file handling module should look for entries in castle. This can be done using intermediate file which will join VFS and Castle modules together.

Castle organizes data into *vertrees* – short for version trees. Each vertree maintains a distinct key-value pairs so data operations on one vertree will not in any way affect data in another vertree[11].Castle uses Doubling array as storage technique, which is similar to LSM and COLA (Cache oblivious lookahead array). Users interact i.e. insert data and query data on vertrees with the help of collections. Collections are views of versions.

## IV. STRUCTURE USED

We have implemented a data structure which would contain filename, D-entry and its struct inode. Each file is given a unique inode number; struct inode has field inode number. With the help of iget(), we get the corresponding inode. Each entry in the table is ordered by a key consisting of the filename string of its directory and structure of its inode.

```
typedef struct ext4_mdata_cache
{
  const char *data;
  struct inode *inode_ptr;
  struct dentry *dentry;
}ext4_mdata_cache;
```

Fig. 6. Data Structure used in Metacache.

At the time of searching the metadata corresponding entry of the file in ext4_mdata_cache is retrieved which ultimately retrieves the inode.

## V. EVALUATION

Currently we have created prototype of MetaCache using simple B-Tree to store metadata. So, when any query comes to VFS regarding access metadata and VFS fails to give the entry, then our prototype will give the corresponding entry. Btree is created in kernel space .These are the results of open system call evaluation between normal module and our module

We have formatted pen drive with Ext4 filesystem & created bunch of files with data. This evaluation is carried out using Strace shell command. :

| % TIME | SECONDS | USECS/CALL | CALLS | ERRORS | SYSCALL |
|---|---|---|---|---|---|
| 100.00 | 0.000226 | 28 | 8 | | MMAP2 |
| 0.00 | 0.000000 | 0 | 1 | | READ |
| 0.00 | 0.000000 | 0 | 7 | | OPEN |
| 0.00 | 0.000000 | 0 | 5 | | CLOSE |
| 100.00 | 0.000226 | | 54 | 5 | TOTAL |

Fig 7: Evaluation of Normal Ext4System performance

```
% TIME   SECONDS  USECS/CALL   CALLS   ERRORS  SYSCALL
------  --------  ----------  -------  ------  -------
100.00  0.000185          23        8          MMAP2
  0.00  0.000000           0        1          READ
  0.00  0.000000           0        7          OPEN
  0.00  0.000000           0        5          CLOSE
    .         .            .        .              .
    .         .            .        .              .
    .         .            .        .              .
------  --------  ----------  -------  ------  -------
100.00  0.000185                    54       5  TOTAL
```

Fig 8: Evaluation of our module (Metacache)

## VI. CONCLUSION

Accessing large collection of small files is always the downside for filesystem performance. Our implementation will improve the performance of Read/Write in Linux File systems and will also improve metadata efficiency.

As LSM tree efficiently stores the data into two trees i.e. Memtable and SStable, most recent data is on the top of tree. This will further increase the performance and therefore will benefit the local filesystem.